\documentclass{article}

\usepackage{graphicx}

\usepackage[dvipsnames]{xcolor}
\usepackage{graphicx} 
\usepackage{biblatex}
\usepackage{caption}
\usepackage{subcaption}
\usepackage{amsmath}
\usepackage{tikz}
\usepackage{multicol}
\usepackage{listings}
\usepackage[htt]{hyphenat}
\usepackage{authblk}

\begin{document}

\title{Algorithm Selection in Short-Range Molecular Dynamics Simulations}

\author[1]{Samuel James Newcome}
\author[1]{Fabio Alexander Gratl}
\author[1]{Manuel Lerchner}
\author[1]{Abdulkadir Pazar}
\author[1]{Manish Kumar Mishra}
\author[1]{Hans-Joachim Bungartz}

\affil[1]{Chair of Scientific Computing in Computer Science, Department of Computer Science,  Technical University of Munich, Boltzmannstr. 3, 85748, Garching bei Muenchen, Germany; samuel.newcome@tum.de}

\date{}

\maketitle

\begin{abstract}
    Numerous algorithms and parallelisations have been developed for short-range particle simulations; however, none are optimally performant for all scenarios. Such a concept led to the prior development of the particle simulation library AutoPas, which implemented many of these algorithms and parallelisations and could select and tune these over the course of the simulation as the scenario changed. Prior works have, however, used only naive approaches to the algorithm selection problem, which can lead to significant overhead from trialling poorly performing algorithmic configurations.

    In this work, we investigate this problem in the case of Molecular Dynamics simulations. We present three algorithm selection strategies: an approach which makes performance predictions from past data, an expert-knowledge fuzzy logic-based approach, and a data-driven random forest-based approach. We demonstrate that these approaches can achieve speedups of up to 4.05 compared to prior approaches and 1.25 compared to a perfect configuration selection without dynamic algorithm selection. In addition, we discuss the practicality of the strategies in comparison to their performance, to highlight the tractability of such solutions.
\end{abstract}

\section{Introduction}

Molecular Dynamics (MD) simulations are computationally intensive simulations in which molecules are modelled as point masses, and force models are used to accelerate the molecules and propagate them through a large number of time steps. They have applications within fields such as thermodynamics \cite{nitzke2025ms2} and bio-membrane simulation \cite{marrink2019computational}. Furthermore, there also exist similar particle simulations, such as Smoothed Particle Hydrodynamics \cite{monaghan2005smoothed} and Discrete Element Method \cite{guo2015discrete}, which use similar algorithms.

Such simulations typically are computationally dominated by inter-particle interactions, such as force calculations. A common optimisation is to introduce a \textit{cutoff} distance, beyond which the interactions are neglected. Simulations with this optimisation are well studied, resulting in the development of numerous algorithms for identifying particle pairs within this cutoff \cite{welling2011efficiency}, as well as parallelisations of those algorithms \cite{tchipev2020algorithmic} \cite{gratl2022n}. However, none of these algorithms is optimal in all scenarios \cite{tchipev2020algorithmic} \cite{gratl2022n} \cite{seckler2021autopas}, and the relative performance of these algorithms can vary on different hardware or with different interaction models \cite{newcome2023towards}.

Such findings motivate the development of the algorithm selection and tuning library AutoPas\footnote{https://github.com/AutoPas/AutoPas}, which aims to select the optimal from a range of over a hundred such algorithmic configurations \cite{gratl2022n}. However, prior works \cite{gratl2022n} \cite{seckler2021autopas} \cite{newcome2023towards} have utilised naive selection strategies, requiring all available configurations to be trialled. Whilst forces calculated with such trialled configurations can still be used to advance the simulation, the worst configurations can perform significantly worse than the best, which can lead to significantly poor performances.

In this work, we present three novel methods that aim to avoid trialling poor configurations. The \textbf{Predictive} strategy uses past performances of a configuration within a simulation to predict how it will perform in later timesteps and avoid trialling those predicted to perform unacceptably below optimum. The \textbf{Expert} strategy uses fuzzy logic rules to describe in which scenarios a particular configuration performs well or poorly, and only trialling configurations which the rules suggest are suited to the scenario. The \textbf{Random Forest} strategy uses a random forest learned on performance data gathered by trialling configurations on artificial simulations.

We present the performances of the selection strategies with three varied Molecular Dynamics simulations, and we particularly highlight the practicalities of such methods applied to some arbitrary particle simulation, which are critical in order to make the use of algorithm selection in Molecular Dynamics tractable.

\section{Background} \label{sec:background}

In this section, we will begin with a brief introduction to Molecular Dynamics simulations before describing AutoPas, its algorithms, parallelisations, and other selection choices, and its selection procedure.

\subsection{Molecular Dynamics}

In Molecular Dynamics Simulations, molecules are simulated as point bodies moving according to Newton's equations of motion, propagated with an integrator such as the Störmer-Verlet method. A simple, commonly used force for uncharged molecules can be derived from the Lennard-Jones potential. Through this, the force acting on each molecule is the sum of the forces acting between that molecule and all other molecules. As such, the force calculation is $O(N^2)$, where $N$ is the number of molecules, whereas the position and velocity updates are only $O(N)$. As such, optimisations of molecular dynamics simulations typically involve optimising the force calculation.

As already discussed, one can introduce a cutoff $r_c$, beyond which the force contributions between a pair of molecules are neglected. For choices of $r_c$ where the number of molecules within a cutoff is significantly fewer than the total, this reduces the force calculation cost to $O(N)$. We will refer to particle pairs that are within $r_c$ as \textit{neighbours}.

A further common optimisation is to use Newton's third law (N3L) to calculate the force between a molecule pair only once and apply that force to both molecules in the pair. Whilst this reduces the number of force calculations in half, writing forces to two molecules at once requires greater protection against race conditions.

For simulations that use the cutoff criterion, a variety of methods exist to calculate the forces efficiently.A \textit{neighbour identification algorithm} (NIA) is employed to reduce the number of distance calculations to $O(N)$\footnote{Assuming that the particles are not extremely heterogeneous.} by identifying, for each molecule, small subsets of the molecules in the simulation which could be a potential neighbour. As force calculations are expensive even with a suitable NIA, parallelisation and data layout considerations are critical to achieving good performance.

\subsection{AutoPas} \label{sec:background.autopas}

There is, however, no NIA, parallelisation, or data layout that is optimal in every scenario, and the best can vary with different densities, homogenities, molecular model, and compute architecture \cite{gratl2022n} \cite{newcome2023towards}. 

This motivates AutoPas, a particle simulation library which provides a black-box particle container to the developer of an arbitrary cutoff-based particle simulator. It contains a variety of internal particle containers, which implement neighbour identification algorithms, several shared-memory parallel traversals, and different data layouts. AutoPas aims to dynamically select the optimal of these algorithmic configurations over the course of the simulation, as well as tune them. As the focus of this work is on the selection strategies, we will provide only a brief introduction to the particle containers, traversals, and other choices available in AutoPas. For a deeper discussion of these, the reader is referred to Gratl et al. \cite{gratl2022n}. 

\subsubsection{Particle Containers} 

\textit{Particle Containers} implement NIAs in AutoPas, which are spatially-aware data structures that store the particles with nearby particles and/or track potentially neighbouring particles. 

The \textit{Linked Cells} (\texttt{LC}) container does this by dividing the domain into cells and binning the molecules into them. Potentially neighbouring particles can, therefore, be determined from neighbouring cells, which are at least partially within $r_c$ of each other. Smaller cells require fewer redundant distance calculations at the cost of more complex memory access patterns.

By default, AutoPas uses cells of dimension $r_c$,  extended by a small skin $\Delta_s$\footnote{The reasoning for this is described by Gratl et al. \cite{gratl2022n}.}, scaled by a \textit{Cell Size Factor} (CSF), and then further extended slightly to perfectly fit the simulation domain. In this work, we will consider a CSF of 0.5 and 1 as possible candidates for selection.

The \textit{Verlet Lists} (\texttt{VL}) container tracks, for each molecule, a list of potential neighbours created by determining which other particles are within an extended sphere of $r_c+\Delta_s$ of the molecule. The lists are rebuilt every $R$ iteration, chosen to guarantee that no particle outside of the extended sphere can become a neighbour without a rebuild occurring.

The other containers used in this paper as candidates for selection but otherwise not used are Verlet List Cells \cite{gratl2022n} and Pairwise Verlet Lists \cite{gonnet2012pairwise}.

\subsubsection{Parallel Traversals} 

Traversals describe how the particles in the container are traversed efficiently with shared memory parallelisation. A full list of traversals used in this work, along with a rough description of their tradeoffs, is presented in Table \ref{tab:traversals}. These range from traversals that parallelise work per molecule (\texttt{List\_Iter}) to those that statically assign entire regions of the domain to each thread (\texttt{SLI}). 

\begin{table}[]
    \centering
    \begin{tabular}{|c|c|c|c|c|}
    \hline
        Traversal & DLB & Scheduling Overhead & Barriers & N3L  \\
        \hline \hline
        \texttt{List\_Iter} & Very good & Very high & 1 & No  \\
        \hline
        \texttt{C01} & Good & High & 1 & Only within cells  \\
        \hline
        \texttt{C08} & Good & High & 8 & Yes  \\
        \hline
        \texttt{C18} & Medium-Good & Medium-High & 18 & Yes\\
        \hline
        \texttt{C04\_HCP} & Medium & Medium & 4 & Yes  \\
        \hline
        \texttt{C04} & Medium-Poor & Low & 4 & Yes  \\
        \hline
        \texttt{SLI\_C02} & Poor & Low & 2 & Yes  \\
        \hline
        \texttt{SLI} & None & None & 1\footnote{The \texttt{SLI} has only one barrier but features a lock per thread.} & Yes  \\
        \hline
    \end{tabular}
    \caption{An overview of the traversals used in this paper, and how well they handle dynamic load balancing (DLB) and scheduling overhead, how many barriers they have, and if they can use Newton's 3\textsuperscript{rd} Law optimisation (N3L). Descriptions of dynamic load balancing and scheduling overhead are simplifications intended only to provide the reader with a rough understanding of the behaviour of the traversals. \texttt{List\_Iter} is only used with Verlet Lists. \texttt{C04\_HCP} and \texttt{C04} are only used with Linked Cells and are not compatible with CSF 0.5. The other traversals are only used with Linked Cells, Verlet List Cells, and Pairwise Verlet Lists. }
    \label{tab:traversals}
\end{table}

\subsubsection{Data Layouts and SIMD}

AutoPas can dynamically switch between the Array-of-Structures (\texttt{AoS}) and Structure-of-Arrays (\texttt{SoA}) data layouts. The latter can take greater advantage of SIMD vectorisation, but where there would be minimal or no benefit from vectorisation, the former provides a faster memory access pattern.

\subsubsection{Distributed Memory Parallelism}

AutoPas is designed to provide MPI-rank level Particle Containers and traversals. Distributed Memory Parallelism is intended to be implemented by the simulator developer. A common method used in this paper is to decompose the domain into a number of subdomains, each assigned to an MPI rank. Molecules crossing the subdomain boundaries need to be sent to the rank of the neighbouring subdomain, and molecules near the boundaries need to also be communicated to the rank on the other side of the boundary so that that rank can calculate and apply the cross-boundary forces. The consequence of this, for AutoPas, is that different regions of the domain (corresponding to different ranks) can use different algorithmic configurations \cite{seckler2021autopas}.

\subsubsection{Algorithm Selection}

AutoPas' algorithm selection process consists of a series of \textit{tuning phases} that occur after a fixed number of timesteps. During each tuning phase, all algorithmic configurations are trialled for a small number of iterations and a weighted average of time spent building the container and time spent calculating forces is taken. The configuration with the smallest weighted average is then used until the following tuning phase. 

Naively trialling all algorithms is referred to as a Full Search.

In this paper, we shall consider the cross-product of all previously discussed configuration options, an alternative \texttt{SoA} variant for \texttt{LC-C01} and \texttt{LC-C04}, as well as N3L enabled and disabled, as possible candidates for selection (excluding incompatible configurations), leading to 116 possible configurations.

\subsection{Fuzzy Logic} \label{sec:fuzzy}

Fuzzy logic provides a mathematical framework for describing ``approximate, rather than exact, modes of reasoning'' \cite{zadeh1988fuzzy}, which is particularly useful for describing logical statements which have some degree of uncertainty in their scope. 

Inputs and outputs are defined by \textit{linguistic terms}, such as ``Low'' and ``High'', which represent \textit{linguistic variables} such as ``Density'' or ``Suitability''. The fuzzy logic system is defined by logical statements of the form \textit{IF antecedent THEN consequent}, where the antecedent is one of these input linguistic terms, or boolean operations upon one or more of them, and the consequent is an output linguistic term. Thus, a possible rule could be \texttt{IF Density == Low THEN Suitability = High}.

The system is evaluated by considering the degree of membership of the input to each linguistic term and evaluating the rules, resulting in a ``fuzzy set'' of degrees of membership to the possible output values. A suitable ``defuzzification'' method, such as taking the centre-of-gravity of this set (as shall be sued in this work), reduces this set to a single ``crisp'' value.

For a more detailed discussion of fuzzy logic, the reader is referred to Zadeh \cite{zadeh1988fuzzy}.

\subsection{Random Forests}

Decision trees are a common machine learning model used for classification tasks, in which a tree of decisions, based on some model input, leads to a single classification decision. Random forests are collections of decision trees that use some kind of splitting of input data so that each tree can make different classification decisions. Each tree ``votes'' for its decided class, and the most popular classification is then the overall chosen classification. This avoids overfitting that might occur with any one tree \cite{breiman2001random}.

\section{Related Work} \label{sec:related_work}

Dynamic Algorithm Selection within High-Performance Scientific Computing is a largely underinvestigated topic.

Armstrong et al. \cite{armstrong2006dynamic} investigated a very similar dynamic algorithm selection problem also applied to short-range particle simulations. They simulated a sparse system of molecules which is contracted into a dense cluster and selected one of two simple algorithms. They used a temporal difference reinforcement learning agent and a linear regression model to predict the optimal method.

Mohammed et al. \cite{mohammed2022automated} investigated the dynamic selection of OpenMP load-scheduling algorithms out of a range of 12 algorithms, which they could rank in one dimension in terms of their scheduling overhead and load balancing capacity. Within this, they considered a random search method, an exhaustive search method, and an expert search method. The expert search method uses expert-written Fuzzy Logic systems that take the loop execution time and a load imbalance metric (or the change in these) as input and output a defuzzified crisp value which maps to an algorithm or a change up or down the ranking. The core concept of this method forms the inspiration for the expert-knowledge Fuzzy Logic system presented later in this work.

Stylianou and Weiland \cite{stylianou2023optimizing} investigated the selection of one of 6 different sparse matrix storage formats using a decision tree and random forest-based selection method. Unlike the previously discussed works, they tested their work on a dataset of independent matrices rather than a problem that evolves over time, and as such, their selection methods are solely based on features from the matrices, such as the average number of non-zeros per row.

While the first of the above works is directly applicable to AutoPas, the latter two are not. As such, whilst these latter two serve as inspiration for an expert-knowledge fuzzy-logic-based approach and a random forest approach, respectively, the ideas must be non-trivially redeveloped into a form which is applicable to AutoPas. Unlike all of the above works, we will consider the much larger number of 116 possible algorithmic configurations.

\section{Live Simulation Statistics} \label{sec:live_sim_stats}

In order to implement expert-knowledge fuzzy logic or data-driven random forest approaches, statistics need to be extracted from the simulation, which such approaches can use as inputs. Whilst every particle's position relative to all others has some impact on the optimality of a configuration, such statistics would be too high dimensional to be practically usable. Instead, particles are binned, and a number of statistics based on the number of particles per bin are produced: the mean, standard deviation relative to the mean, median, and maximum. We use cell bins, which mimic a CSF 1 cell, independent of the actual configuration being used. In addition, the number of cell bins, the number of empty cell bins, the thread count, and the skin are collected. Such statistics can be produced with $O(N)$. 

\section{AutoPas' Selection Strategies} \label{sec:selection_strats}

In this section, we will describe three novel algorithm selection methods for AutoPas: the \textbf{Predictive}, \textbf{Expert}, and \textbf{Random Forest} methods.

\subsection{The Predictive Strategy}

In the first two tuning phases, each configuration is trialled. After this, at the start of each tuning phase, a linear model for each configuration, fitted to the last two data points of that configuration, is used to predict the performance of that configuration in that tuning phase. Only the configurations expected to perform within a relative threshold of the expected optimum are trialled.

To avoid configurations that previously performed poorly never being trialled again, even if the simulation has changed significantly, we retrial configurations after a number of tuning phases where they have not been trialled, even if they are expected to perform worse than the threshold. 

In addition, we further apply a slow configuration filter such that configurations that perform more than a relative threshold worse than the best configurations are blacklisted so that they are never tested again during this simulation.

In this work, we will trial configurations with expected performances within $2\times$ the expected optimal, retrial configurations that have not been trialled for approximately a quarter of the simulation, and blacklist configurations that perform $10\times$ worse than the optimal. 

\subsubsection{Practicality of Method}

The predictive tuning strategy's simplicity is a clear benefit compared to later methods. No data is required outside of the data generated within the method itself, and, to a degree, no expert knowledge is required to use it. The method could benefit from optimising the parameters for a particular simulation; for example, more dramatically changing simulations might benefit from more frequent trialling of configurations.

\subsection{The Expert Strategy}

A fuzzy-logic-based system was designed to map the simulation statistics using expert-written fuzzy-logic rules to choose an optimal algorithmic configuration. Fuzzy-logic is a good choice for such a system due to the ``fuzzy'' nature of an expert explaining why certain configurations perform well in some situations: we might describe a configuration as being suitable because of a ``high'' simulation density but could not provide an exact density at which the configuration is best.


\subsubsection{Practicality of Method}

The clear downside of this method is the requirement for an expert to develop such rules. The influence of simulation statistics upon the optimality of configurations is nonlinear, and therefore, even ignoring the influence of the hardware and force model, developing a model that fully describes why a given configuration is suitable for a particular scenario would require significant effort.

Furthermore, even if perfect rules for a given hardware and force model could be developed, Newcome et al. \cite{newcome2023towards} showed that the relative optimality of configurations varies between hardware and force models. And even if the rules could be adapted to different hardware, AutoPas is designed to be used by potentially non-expert developers of arbitrary particle simulators with any kind of interaction model. 

Finally, even ignoring the above points, as AutoPas itself is developed, with new configurations and improvements upon existing configurations, such rules would have to be adapted.

Whilst the above points suggest the infeasibility of a general set of expert fuzzy-logic rules that could be universally applied, the strategy could be used to design simple rules that merely indicate ``families'' of potentially optimal algorithms, explicitly tailored to the use case.

\subsubsection{Designing suitable rules}

As such, a set of rules was developed specifically for the example simulations below. These were designed by running the example simulations with regular naive, full searches and comparing the live simulation statistics with when certain configurations are chosen.

Whilst this methodology requires data from existing simulation runs to inform the expert's decision, it is a valid strategy when the expert can use such data-informed rules again on further similar simulations. Such further similar simulations could include conducting several experiments with a range of parameters \cite{maeda2003solid}, when using the replica-exchange molecular dynamics method \cite{sugita1999replica}, or through using checkpoints to continue a simulation. 

The mean number of particles per cell-bin was found to significantly indicate ``families'' of potentially optimal algorithms, and so the fuzzy logic system used this variable alone to describe the suitability of a subset of 11 of the total 116 configurations, that were found to at some point be the optimal, in terms of ``Good'', ``Okay'', or ``Bad'' based on this mean.

We should note that other works \cite{seckler2021autopas} \cite{newcome2023towards} have found other configurations to be optimal, and therefore, assuming the other 105 configurations should always be dismissed would be naive.

\subsection{The Random Forest Strategy}

To implement the random forest strategy, the Python/C API is used to call SciPy's random forest model from C++ by passing the live simulation data as a JSON object to Python, which then passes this as an input to a random forest model that suggests a configuration which gets returned to C++ as a JSON object. The purpose of using Python was due to this being a typical language for machine learning, which could aid the future development of more advanced machine learning models through the easy use of existing Python models and libraries and the ease of collaboration with machine learning experts who may be less able C++ developers.

\subsubsection{Practicality of Method}

Such a data-driven method is far easier to use than the Expert strategy. Even though the relationship between simulation and optimal algorithmic configuration is hardware and interaction model dependent, if a user has a set of simulation scenarios, they can trial AutoPas' configurations upon that set, thus generating data relating the live simulation statistics for those scenarios and the performances of the configurations for their particular hardware and simulator. This avoids much of the difficulties mentioned previously for expert-knowledge methods, as if the set of scenarios is already provided, trialling the configurations and training a model requires only a minor amount of effort. 

The key issue, in terms of the practicality of any machine learning approach, comes from generating a set of simulation scenarios upon which the configurations can be trialled. Firstly, generating data for mid-simulation scenarios would require many timesteps to reach and, therefore, require high computational costs or uploading large checkpoints for such scenarios from some online repository, which could make data generation difficult. Secondly, whilst requiring no expert input for the relations between live simulation statistics and the performance of configurations, the simulation scenarios need to relate to the simulations which the user intends to optimise with an ML-based strategy. For example, with Lennard-Jones forces, the training data should include densities up to the maximum density achievable with the force parameters the user wishes to use. However, generating data for even denser simulations would be a waste of computational resources. Therefore, some expert decisions regarding the dataset still need to be made.

\subsubsection{The Dataset}

To avoid the first of these issues, we generated scenarios using simple random uniform and Gaussian distributions as well as tight hexagonal-packed grid structures and spheres carved from regular grid structures. Such scenarios are likely unrealistic based on the exact molecule placement (molecules too close together) but are representative at the cell-bin level. 

To address the second issue, these ``fake'' scenarios were designed to mimic a variety of different ``real'' scenarios, ultimately with the example simulations in mind. These were
\begin{itemize}
    \item Uniform distributions of particles on a $20\times20\times20$ domain with numbers of particles ranging from 100 to 204000, increasing in factors of two. 
    \item 10 to 160 individual Gaussian distributions, increasing in factors of two, where each Gaussian distribution has a standard deviation of 2 in all dimensions and is randomly uniformly distributed throughout a $16\times16\times16$ domain, with domain borders that bound the resulting distribution with a buffer of 0.5 on all faces. 
    \item A $40\times40\times40$ domain with 4000 randomly uniformly distributed molecules with a tight hexagonal-packed cuboid in the middle with a particle spacing of 1.225 and a volume of $x\time40\times40$, where $x$ is varied between 5 and 20 in intervals of 5.
    \item A sphere in the centre of a $200\times200\times200$ domain with radii from 5 to 20 in intervals of 5 and particle spacings from 0.5 to 2 in intervals of 0.5.
    \item An empty $40\times40\times40$ domain.
\end{itemize}

For all scenarios, skins between 0.1 and 0.5, in intervals of 0.1, and numbers of threads between 6 and 36, in intervals of 6, were trialled. Each configuration was trialled for 10 iterations, one of which featured a particle container build. For all scenarios, a cutoff of 3 was used. 

\subsubsection{The Random Forest Model}

A Random Forest Classifier with 100 estimators was used. As input features, it used all the statistics described in Section \ref{sec:live_sim_stats}. The Random Forest Classifier model of \texttt{scikit-learn 0.24.2} was used.

\section{Experimental Setup} \label{sec:exp_setup}

We tested our tuning strategies on HSUper\footnote{https://portal.hpc.hsu-hh.de/documentation/hsuper/}, whose compute nodes feature 256GB of RAM, 2 Intel Icelake sockets each with an Intel(R) Xeon (R) Platinum 8360Y processor with 36 cores. The data used to train and develop the tuning strategies was also produced on the same nodes of HSUper. The full details of the versions of AutoPas, the compiler, MPI, Python and its modules can be found in Appendix \ref{app:versions}.

We selected three scenarios to demonstrate our strategies on three scenarios, ranging from the sparse no-MPI Heating Sphere experiment to the dense 40 MPI-rank Rayleigh Taylor scenario. Between these scenarios, there is minimal overlap in the optimal algorithm configuration.

In all experiments, a cutoff of 3 is used.

\subsection{Heating Sphere}

The heating sphere experiment, a small sphere of 5497 regular Lennard-Jones molecules, is placed in the centre of a large $200\times200\times200$ domain with reflective boundaries. This is simulated for $1.5\times10^{5}$ iterations with $\delta t = 0.0001$. A skin of 0.5 is used with a rebuild interval of 10. The simulation starts with a temperature of 0.1, and, through applying a simple velocity scaling thermostat in intervals of 100 iterations with small maximum changes in temperature of 0.1, the temperature is slowly increased to 100. The simulation starts with this small ball closely bound together, and over time, the molecules escape the sphere, sparsely filling the domain. The simulation was run on one compute node with 1 MPI rank with 36 hardware threads. The optimal algorithmic configuration begins with \texttt{VL-List\_Iter-NoN3L-AoS} and then switches to \texttt{LC-C04-N3L-AoS-CSF1}.

\subsection{Exploding Liquid}

In the exploding liquid experiment, a thick, tight hexagonal packed $120\times18\times120$ slice of 370440 regular Lennard-Jones molecules are placed in the centre of a $120\times480\times120$ domain, with periodic boundaries. On either side of the slice are 100 random uniformly distributed molecules. A skin of 0.6 is used with a rebuild interval of 10. The thick slice explodes outwards, along the $y$-axis, with two dense ``waves'' of particles in each direction, and many particles remain behind the waves in small or medium clusters. The domain is split into six sub-domains along the $y$-axis, each with its own AutoPas container that makes independent algorithmic choices. The simulation was run on two compute nodes with 6 MPI ranks, each with 24 hardware threads. Ranks whose regions cover a ``wave'' have \texttt{LC-C04\_HCP-N3L-SoA-CSF1} as the optimal configuration, those where the ``wave'' has not reached yet have \texttt{VL-List\_Iter-NoN3L-AoS}, and those after the ``wave'' have left typically have \texttt{LC-C04\_HCP-N3L-SoA-CSF1} or \texttt{LC-C08-N3L-SoA-CSF1}. Occasionally, other configurations are optimal.

\subsection{Rayleigh-Taylor}

In the rayleigh-taylor instability experiment, which is a classic experiment within particle simulation, a layer of larger lighter molecules is placed under a layer of smaller heavier molecules, with the molecule layers mixing with the larger molecules rising to the top and the smaller molecules sinking to the bottom. We used, for the bottom layer, 109461 regular Lennard-Jones molecules placed in a tight hexagonally packed cube of dimensions $60\times60\time30$ with a particle spacing of 1.122. For the top layer, we used 719200 half-size, double-mass Lennard-Jones molecules in a tight hexagonally packed cube of dimensions $60\times60\time30$ with a particle spacing of 0.6. A gravitational force of -12.44 is applied downwards. A skin of 0.3 is used with a rebuild frequency of 30. The simulation was run on five compute nodes with 40 MPI ranks, each with nine hardware threads. Ranks whose regions contain lots of half-size molecules have \texttt{LC-C08-N3L-SoA-CSF0.5} or \texttt{LC-C01-NoN3L-SoA-CSF0.5} as an optimal configuration, those with more regular molecule typically have the \texttt{LC-C08-N3L-SoA-CSF1} as an optimal configuration. Many other configurations may also be optimal, however these are the most prominent configuration.

\section{Results \& Analysis} \label{sec:results}

\begin{table}[]
    \centering
    \begin{tabular}{|c||c|c|c|}
        \hline
        Configuration & Heating Sphere & Exploding Liquid & Rayleigh-Taylor \\
        \hline
        \hline
        \texttt{LC-C04-N3L-AoS-CSF1} & $2.754\times10^2$ & $8.875\times10^3$ & - \\
        \hline
        \texttt{LC-C04\_HCP-N3L-SoA-CSF1} & $9.193\times10^2$ & $2.947\times10^3$ & $7.792\times10^5$ \\
        \hline
        \texttt{LC-C08-N3L-SoA-CSF0.5} & $2.307\times10^4$ & $2.370\times10^4$ & $4.474\times10^5$ \\
        \hline
        
    \end{tabular}
    \caption{Total time spent calculating forces for all threads for each experiment for the algorithmic configurations that were the most optimal single configuration for each experiment described in Section \ref{sec:exp_setup}. Note that the \texttt{LC-C04-N3L-AoS-CSF1} with Rayleigh-Taylor experiment timed out upon reaching the maximum wall time allowed on HSUper - about 7 times the wall time that \texttt{LC-C08-N3L-SoA-CSF0.5} took and only achieving approximately three-quarters of the total iterations.}
    \label{tab:best_single_configs}
\end{table}

Reinforcing the previously discussed concept that there is no optimal algorithmic configuration in all scenarios, we found that there was no single best algorithmic configuration for all three experiments, and the best for one experiment could perform terribly on another. In Table \ref{tab:best_single_configs}, the best single algorithmic configuration in each experiment is compared with those of the other experiments. We see that the best configuration for the Heating Sphere and Rayleigh-Taylor perform significantly worse than each other in the other experiment. \texttt{LC-C04\_HCP-N3L-SoA-CSF1} generalises better to all three experiments but still performs a significant 3.338 times slower than the optimal in the Heating Sphere experiment.

\begin{figure}
\begin{subfigure}[b]{.3\textwidth}
\centering
\includegraphics[width=\textwidth]{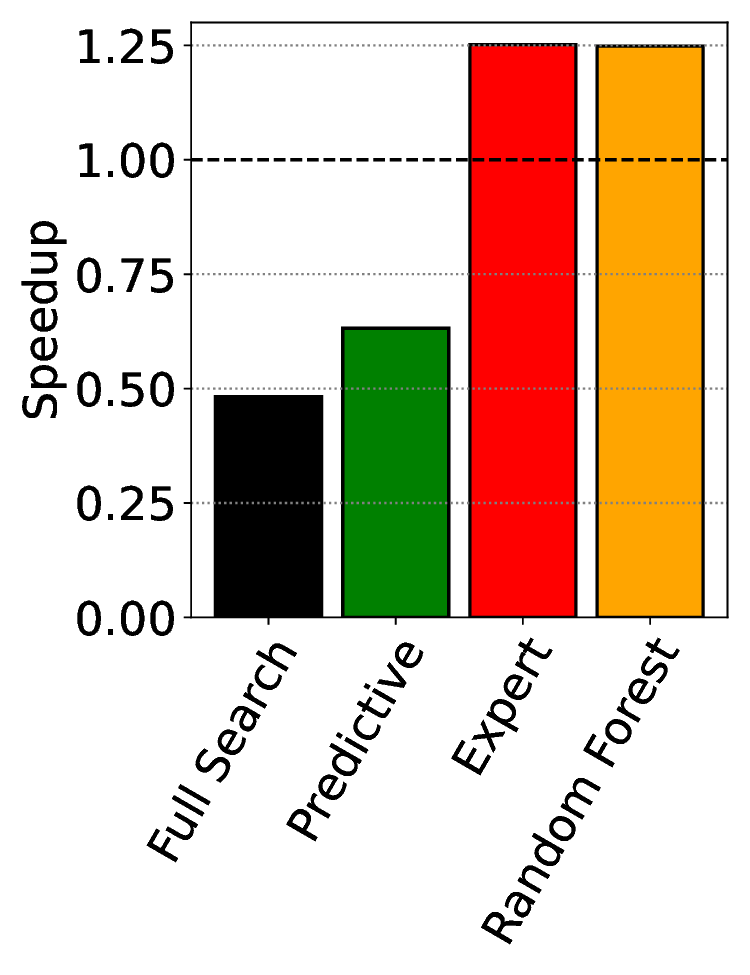}
\caption{Heating Sphere.}
\label{fig:results:HS}
\end{subfigure}
\begin{subfigure}[b]{.3\textwidth}
\centering
\includegraphics[width=\textwidth]{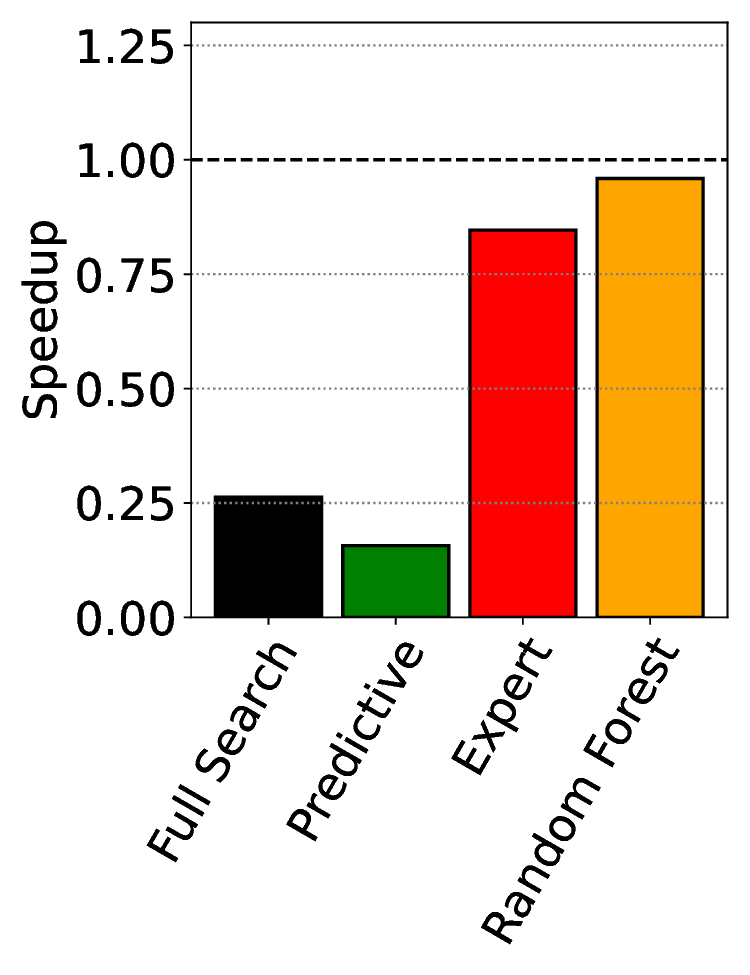}
\caption{Exploding Liquid.}
\label{fig:results:EL}
\end{subfigure}
\begin{subfigure}[b]{.3\textwidth}
\centering
\includegraphics[width=\textwidth]{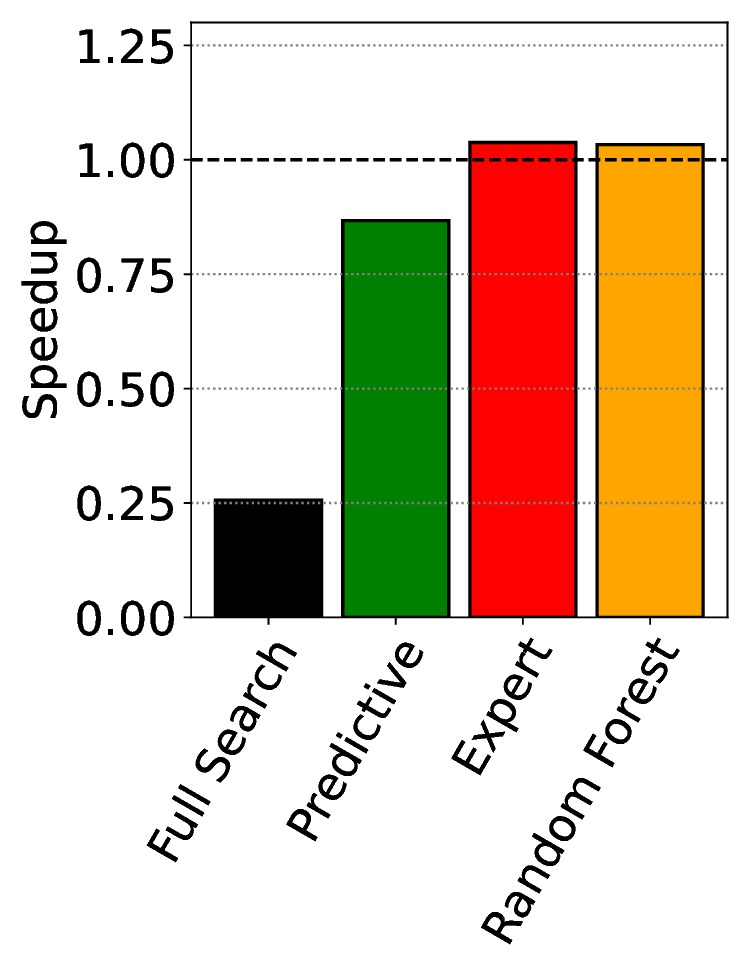}
\caption{Rayleigh Taylor.}
\label{fig:results:RT}
\end{subfigure}

    \caption{Comparison of the total time spent calculating forces for each thread across the entire simulation for each tuning strategy relative to the optimal single configuration for each experiment. These are (a) \texttt{LC-C04-N3L-AoS-CSF1}, (b) \texttt{LC-C04\_HCP-N3L-SoA-CSF1}, (c) \texttt{LC-C08-N3L-SoA-CSF0.5}.}
    \label{fig:results}
\end{figure}

In Figure \ref{fig:results}, the speedups of each strategy tested are plotted relative to the single best algorithmic configuration in each experiment. 

Note that we have different degrees of success for each strategy. A slowdown seen with Predictive or Random Forest is not a failure of the method, as a user without prior knowledge of which single algorithmic configuration is the best will still benefit from achieving a close-to-the-best performance. Meanwhile, the suggested methodology for developing the Expert strategy implies that this best single configuration is already known, and therefore, this method is only of value if it can achieve a speedup.

The Predictive strategy achieves limited success. The Exploding Liquid experiment features dramatic changes in the computational profile of each MPI rank, which would make local linear models of configuration performances often unsuitable, and therefore the worse performance of Predictive Tuning compared to multiple naive Full Searches is expected. In the Heating Sphere experiment, \texttt{LC-C04-N3L-AoS-CSF1} is, however, blacklisted. This appears to be due to some internal inefficiencies that arise in the second tuning phase but do not appear elsewhere in this work and are not yet understood.  

The Expert and Random Forest strategies perform similarly, with Random Forest performing slightly better in the Exploding Liquid scenario. In the Heating Sphere experiment, both strategies achieve speedups of 1.25, correctly following the switch in the optimal configuration. This represents the significant potential of the methods and dynamic algorithm selection more generally. In the Exploding Liquid experiment, however, the methods achieve ``speedups'' of 0.85 and 0.89, respectively and in the Rayleigh-Taylor experiment, they achieve 1.04 and 1.03, respectively.

In these experiments, whilst the optimal algorithmic configuration does change between the MPI ranks and over time in both experiments, the MPI ranks with the highest load typically have \texttt{LC-C04\_HCP-N3L-SoA-CSF1} (in Exploding Liquid) or \texttt{LC-C08-N3L-SoA-CSF0.5} (in Rayleigh-Taylor) as the optimal configuration, even if the MPI-rank with this highest load changes. This explains the inability of the Expert and Random Forest strategies to achieve (significant) speedups over picking the single best configuration and suggests that for such speedups to be possible, the optimal configuration in the highest workload MPI rank has to change.

\section{Conclusion and Outlook}

Of the three methods presented in this work, the Random Forest strategy appears the most successful -- balancing performance with ease of use -- and is able to achieve moderate speedups of 1.25 compared to simply picking the optimal configuration and 4.05 over the previous naive approach. Whilst it is expected that the Expert strategy could be improved to match or potentially beat the Random Forest strategy, this is not a tractable solution to the problem and is unsustainable. The Predictive strategy struggles significantly, particularly in dramatically changing scenarios; it is an easier approach to use, and the further development of such an accessible, data-free approach could be valuable in some scenarios.

As previously mentioned, a key issue with such data-driven models is the need for good data, the generation of which is itself an expert decision. A more accessible solution could be to combine online learning with an understanding of confidence: in scenarios where the model has not been trained on similar data, the model becomes unconfident in its suggestions, triggering exploration of the search space and thus online learning, and therefore better performance in further similar simulation runs. Thus, the user generates data relevant to the simulations they want to optimise.

\section{Acknowledgements}

Computational resources (HPC-cluster HSUper) have been provided by the project hpc.bw, funded by dtec.bw—Digitalization and Technology Research Center of the Bundeswehr. dtec.bw is funded by the European Union—
NextGenerationEU. We acknowledge financial support by dtec.bw—Digitalization and Technology Research Center of the Bundeswehr. dtec.bw is funded by the European Union—NextGenerationEU.

The authors would like to thank Amartya Das Sharma and Ruben Horn, of Helmut-Schmidt-University, for their invaluable assistance and advice with the Python-C interface on HSUper. The authors would like to express their deep thanks to all contributors of the open-source AutoPas project.

\appendix

\section{Software Versions} \label{app:versions}

Both training data and results were produced using AutoPas' example simulation, md-flexible, with commit ID \texttt{a3b2acc}. AutoPas and md-flexible share a repository, thus it is the same commit ID for both. These were compiled using GCC 13.2.0 with Intel OneAPI MPI version 2021.12.1. For the random forest tuning strategy, Python 3.6.8 was used together with numpy 1.19.5, pandas 1.1.5, scikit-learn 0.24.2 and scipy 1.5.4. 

\section{Input Files and Data}

All input files, outputs, job scripts for training data and the experiments, as well as models and fuzzy rule files, can be found at\ https://github.com/SamNewcome/Algorithm-Selection-in-Short-Range-Molecular-Dynamics-Simulations.

\printbibliography

\end{document}